\documentclass[onecolumn,10pt]{article}
\usepackage{graphicx}
\usepackage[latin1]{inputenc}

\usepackage[T1]{fontenc}
\usepackage[francais]{babel}

\begin{document}

\begin{center}
{\bf
{\Large The Sun Asphericities: Astrophysical Relevance } \\
}
\end{center}

\vspace{0.8truecm}
\centerline{Jean-Pierre ROZELOT$^1$, Sophie PIREAUX$^2$, Sandrine LEFEBVRE$^3$}
\centerline{$^{1}$ Observatoire de la Côte d'Azur (OCA)}
\centerline{Département GEMINI - 06130 GRASSE, France}
\centerline{$^{2}$ \small Observatoire Midi-Pyrénées (OMP, UMR 5562), 31400 Toulouse, France}
\centerline{$^{3}$ \small University of California UCLA, Los Angeles, CA 90096-562, USA, on leave from OCA}

\medskip
\centerline{\footnotesize e-mail: jean-pierre.rozelot@obs-azur.fr}
%
\setcounter{page}{1}
\vspace{1truecm}
\bigskip

{\footnotesize Of all the fundamental parameters of the Sun (diameter, mass, temperature...), 
the gravitational multipole moments (of degree $l$ and order $m$) that determine the solar 
moments of inertia, are still poorly known. However, at the first order (l=2), the quadrupole 
moment is relevant to many astrophysical applications. It indeed contributes to the 
relativistic perihelion advance of planets, together with the post-Newtonian (PN) parameters; 
or to the precession of the orbital plane about the Sun polar axis, the latter being 
unaffected by the purely relativistic PN contribution. 
Hence, a precise knowledge of the quadrupole moment is necessary for accurate orbit 
determination, and alternatively, to obtain constraints on the PN parameters.
Moreover, the successive gravitational multipole moments have a physical meaning: 
they describe deviations from a purely spherical mass distribution. Thus, their precise 
determination gives indications on the solar internal structure. 
Here, we explain why it is difficult to compute these parameters, how to derive the best 
values, and how they will be determined in a near future by means of space experiments.}

{\bf
\centerline{
 }

\centerline{
  \footnotesize general relativity; perihelion advances of planets.}
}

\smallskip
\noindent
\bigskip


\section{Introduction}

{\normalsize The study of the rotation of stars is not trivial. In theory, the problem is 
exceedingly simple and can be formulated as follows. Consider a single star that rotates 
about a fixed direction in space, with an angular velocity $\Omega$. Let us first assume that, 
for $\Omega = 0$, the star is a gaseous body in gravitational equilibrium. The problem is to 
determine the outer shape of the star when the initial sphere is set rotating at an angular 
velocity $\Omega$. Such studies were conducted for the first time by Milne (1923), then fully 
achieved by Chandrasekhar (1933).

The second point which arises is to ask oneself what will happen if $\Omega$ is not constant, 
not only in latitude (differential rotation) but also throughout the body, from the surface 
to the core. Today, astronomers are faced with such problems, not only in the solar case, 
but also for stars. With the advent of sophisticated techniques such as interferometry, one 
is now able to accurately determine the geometrical shape of the free boundary of stars, such 
as Altair or Achernar for which observations of the geometrical enveloppe have been made by 
Belle {\it et al.} (2001), and Domiciano de Souza (2002). But it would be of little or no
interest to observe the geometric shape of a star if one would not be able to infer some 
information on stellar physics. With such an approach, the purpose of theoreticians is 
enumerate all the possible angular velocity distributions (from the center to the surface) 
that are compatible with the observed stellar surface. For stars, Maeder (1999) examined 
the effects of rotation and wrote the equation of the surface with a rotation law which is 
differential, but only in the surface layer.

In other words, the knowledge of the angular velocity distribution from the core to the 
surface, together with the knowledge of the density function (related to the pressure 
function), completely determine the outer shape of the stars. Different techniques exist to 
observe such a figure. Once accurately determined, one would be able to go back to the 
physical properties of the body. This approach is called {\it ``Theory of Figures''}.

The Theory of Figures has been widely used in geophysics and is still used in specific cases, 
such as for the planet Mars, with an incredible accuracy ($J_{2~Mars}$ = 1.860718 10$^{-3}$ 
according to Yuan {\it et al.} (2002), from a 75th degree and order model). Since the 
pioneering work of Clairaut (1743), who was the first to compute the flattening of a rotating 
body, and Bruns (1878) who introduced the concept of ``Figure of the Earth'', considerable 
work has been done, mainly by Radau (1885), Wavre (1932), Molodensky (1988), Moritz (1990), 
and many others since then. 
The basic principle of this theory is to determine the outer surface of a rotating body, 
assuming a knowledge of the mass and angular velocity distributions.
In the case of the Earth, the main emphasis was on the determination of a global ellipsoid. 
The theory of hydrostatic equilibrium was considered as the best means for determining the 
terrestrial flattening $f$, until the advent of artificial satellites in 1957. Today, 
considerable accuracy is reached that allows the inverse problem to be solved, that is to say 
to determine anomalies inside the Earth mantle. The Earth external gravitational potential 
is developed in a serie of so-called spherical harmonics of degree $l$ and order $m$. The 
dimensionless coefficients represent the different distortions from a pure sphere. Thanks to 
artificial satellites, they are now computed up to a very high degree 
($J_{2~Earth}$ = 4.84165209 10$^{-4}$ according to the EGM96 tide free model adopted by the 
International Earth Rotation Service in (McCarthy, 2003), from a 360th degree and order model), 
and substantial progresses are made in the knowledge of our planet. 

Curiously, for the Sun, the situation is hardly that of the Earth which prevailed before 1957. 
The flattening of the Sun is still poorly known and we hardly know if there is a variability 
linked to the solar cycle. The Theory of Figures was not developed for the Sun until Rozelot 
{\it et al.} (2001). 
The difference between the Sun and the Earth lies in the fact that, in the latter case, 
the rotation can be considered as solid and the outer shape is mainly due to strong density 
variations inside the body, on which are superimposed tides, oceanic currents and atmospheric 
effects. On the contrary, in the solar case, the density can be considered as nearly 
homogeneous on successive shells, whereas the rotation is strongly dependent on the latitude.
Solar modeling is further complicated by the fact that this differential rotation is not only 
a surface phenomena, as can be seen for example by observing spots or faculae, but 
is anchored in depth, at least down to 0.7 solar radii. The computation of the solar 
gravitational potential may be not too difficult in its formalism, and is nearly the same as 
for the Earth itself. However the computation of the centrifugal potential is very complicated 
and it cannot be reduced to computation of potentials on successive cylinders (or thin zonal 
rings) in which the rotation is taken as uniform. Moreover, we have shown that the centrifugal 
force corresponding to the commonly adopted differential rotation law (a development in cosine 
of the colatitude) does not derive from a potential. Hence, we proposed an alternative 
rotation law (Rozelot and Lefebvre, 2003). The complexity of the rotation 
profile (see for example Eq. \ref{eq.rotde}) indicates that the photospheric shape is highly sensitive to 
the interior structure. Thus, in principle, accurate measurements of the limb shape distortions, 
which we call {\bf ``asphericities''} (i.e. departures from the ``helioid'', the reference 
equilibrium surface of the Sun), combined with an accurate determination of the solar rotation 
provides useful constraints on the internal layers of the Sun (density, shear zones, surface 
circulation of the plasma...).  Fig. \ref{fig:AsphericalSurface} shows such asphericities 
that can be seen at a given spatial resolution.
Alternatively, theoretical upperbounds could be inferred for the flattening which may exclude 
incorrect/biased observations.
\newline
Thus, even if we know that the Solar Theory of Figures is limited due to truncation errors, 
the learning is rich.

In this paper, we will successively focus on: (i) the definition of the solar gravitational 
moments, (ii) how to compute these moments, (iii) their relevance to astrophysics and (iv) 
how to measure such quantities. To conclude, we will show why space missions are needed.}

\begin{figure}[t]
\center
        \includegraphics[width=.7\textwidth]{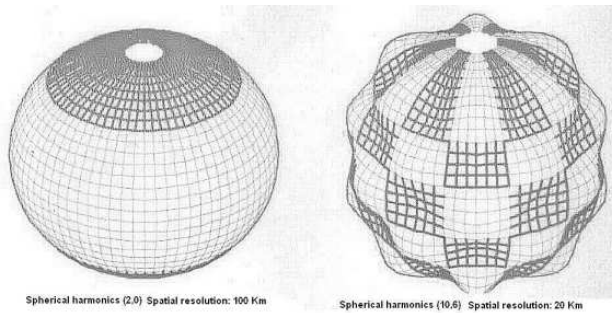}
        \caption{\small Left: 3d sketch of the ($l$= 2, $m$= 0) mode; 
                  the Sun is nearly a spheroid. 
                  Right: 3d sketch of the ($l$= 10, $m$= 6) mode; 
                  distortions from a spherical surface appear, on a very 
                  exagerated scale.}
        \label{fig:AsphericalSurface}
\end{figure}

\section{Definition of the solar gravitational multipole moments}

{\normalsize For an axially symmetric distribution of rotating matter, the outer gravitational 
field can be expressed as:
\begin{equation}\
  \phi_{grav~out}(r,\theta)=-\frac{GM_{\odot}}{r}\left[1-\sum_{n=1}^{\infty}\left(\frac{R_{\odot}}
  {r}\right)^{2n}J_{2n}P_{2n}(\cos\theta)\right]
\label{phi}
\end{equation}
where $G$ is the gravitational constant; $M_{\odot}$ and $R_{\odot}$, the solar mass and radius; 
$J_{2n}$ are the gravitational multiple moments; $P_{2n}$, the Legendre polynomials; 
$r$ and $\theta$, respectively the distance from the Sun centre and the angle to the symmetry 
axis (colatitude).
 
\noindent
The first terms are called
\begin{itemize}
        \item for n = 1, $J_2$ the solar quadrupole moment
        \item for n = 2, $J_4$ the solar octopole moment
        \item for n = 3, $J_6$ the solar dodecapole moment
        \item ...
\end{itemize}
For the Sun, as we will see later, terms of  higher degree are nearly null.
The $J_{n}$ coefficients are dimensionless quantities providing information on how the mass 
and velocity distributions act inside the Sun to finally render the outer visible shape non 
spherical. The amplitude, peak to peak, of such  asphericities does not exceed some 20 
milliarcseconds (abbreviated as ``mas''), as shown by the theory itself (Lefebvre and Rozelot, 
2004) or by observations (Rozelot {\it et al.}, 2003). }

\section{Computation of the solar gravitational moments}

{\normalsize The starting point is to consider the equation of motion for an ideal fluid 
\begin{equation}
        \rho {\vec {\ddot{x}}} = \rho {\vec g} - {\vec \nabla} p 
\label{clairaut}
\end{equation}
where ${\vec {\ddot{x}}}$ is the acceleration of the fluid; $\rho$ is the density; 
${\vec g}$, the gravitational force ; and ${\vec \nabla} p$, the pressure gradient. 
For hydrostatic equilibrium, there is no motion, hence ${\vec {\ddot{x}}}$ = 0. 
For a nonrotating body, the force acting on particles is given by 
${\vec g}=-{\vec \nabla} \phi_{grav} $, 
where $\phi_{grav}$ is the gravitational potential. Hence Eq. \ref{clairaut} reduces to
\footnote{The curl of this equation is ${\vec \nabla} \rho$ {\bf X} ${\vec \nabla} \phi = 0$, 
so that the normals to surfaces of constant $\rho$ and $\phi$ point in the same direction.
As a consequence, the surfaces of constant $\rho$ and $\phi$ coincide. Using the perfect gas 
equation of state (with constant chemical composition), it can be shown that surfaces of 
constant $\rho$, $\phi$, $T$ and $P$ all coincide. This is known as the Von Zeipel theorem: 
any internal source of distortion in the gravitational field at the surface will manifest 
itself as a change of shape in the solar surface layer. Thus, measuring the shape of the 
surface layers is equivalent to measuring surfaces of constant gravitational potential.} 
\begin{equation}
        0 = -\rho {\vec \nabla} \phi_{grav} - {\vec \nabla} p 
\label{clairautdeux}
\end{equation}
When the body is rotating, the force acting on particles is the gradient of the gravity 
potential ($\phi_{total}=\phi_{grav}+\phi_{rotation}$); that is, we must add the centrifugal 
force, so that Eq. \ref{clairautdeux} becomes
\begin{equation}
        {\vec \nabla} p  =  - \rho {\vec \nabla} \phi_{grav} + \Omega^{2}(r,\theta)~\textbf{${\vec s}$}
\label{clairauttrois}
\end{equation}
where $\Omega(r,\theta)$ is the angular velocity and ${\vec s}$ a vector perpendicular to the 
rotation axis directed outwards.

\bigskip
There are two ways to compute the coefficients $J_n$. 

\medskip
\noindent
$\bullet$ The first method solves Eq. \ref{clairauttrois} by considering that asphericities 
are small quantities: the solution is the sum of a spherically symmetric potential (indexed 
by a subscript 0) and a perturbation (nonspherically symmetric, indexed by a subscript 1):
\begin{equation}
  \phi_{grav}(r,\theta)=\phi_{grav~0}(r)+\phi_{grav~1}(r,\theta)
\end{equation}
Such an approach was taken by Goldreich and Schubert (1968), Ulrich and Hawkings (1981), 
Pijpers (1998), Armstrong and Kuhn (1999), Godier and Rozelot (1999) or Roxburgh (2001). 
Differences lie in the way Eq. \ref{clairauttrois} is projected on appropriate coordinates. 
Except for Armstrong and Kuhn who used a series of vector spherical harmonics to avoid 
truncation errors, all the above mentioned authors made a projection on Legendre polynomials 
$P_{2n}(cos\theta)$, so that 
\[ \phi_{grav~1}(r,\theta) = \sum_{n=1}^{\infty}\phi_{grav~1,2n}(r)~P_{2n}(cos\theta) \]
According to this, Eq. \ref{clairaut}, supplemented with Poisson's equation
\begin{equation}
        \nabla^{2}\phi_{grav}=4\pi G \rho       
        \label{poisson}
\end{equation}
and boundary conditions at $r$ = 0, where $\phi_{grav}=0$ or at the surface $r$= $R_{\odot}$ 
where $\phi_{grav}=\phi_{grav~out}$, finally lead to
\begin{equation}\
  J_{2n}=\frac{R_{\odot}}{GM_{\odot}}~\phi_{grav~1,2n}(R_{\odot})
        \label{solution_J2n}
\end{equation}
The function $\phi_{grav~1,2n}$ is the solution to a differential equation requiring the 
knowledge of $\rho (r)$ and $\Omega(r, \theta)$. 
Different models can be used for the density or the rotation law. In general $\rho (r)$ is 
taken either from Richard (1999), Christensen-Dalsgaard {\it et al.} (1996) or Morel (1997). 
The rotation law can be either the standard rotation law (in cosine, see Eq. 
\ref{standard_rotation_law}) or can be derived from 
the analytical law of Kosovichev (1998) or Corbard {\it et al.} (2002) (see Eq. 
\ref{Corbard_Dikpati_analytical_rotation_law}). 
Fig. \ref{corbard} -left- shows solar rotation profiles, from the surface down 
to 0.3 $R_{\odot}$, obtained by a 1.5 dimensional inversion of data from the $MDI$ (Michelson 
Doppler Imager) experiment on board the SOHO mission (Di Mauro, 2003). 
Fig. \ref{corbard} -right- shows the Corbard rotation model deduced from observations of 
the $MDI$ ~$f$-modes between May 1996 and April 2001 (Dikpati {\it et al.}, 2002). 
\newline
A complete expression of $\phi_{grav~1,2}$ and $\phi_{grav~1,4}$ was provided by Armstrong 
and Kuhn (1999), using the standard rotation law.
However, surface plasma observations allow to constrain analytical rotation models. 
The first attempt to derive an analytical rotation law from helioseismic data has been made 
by Kosovichev (1996). Using this law, Godier and Rozelot (1999), then Roxburgh (2001) computed the gravitational 
moments (see Table \ref{momgrav}). The discrepancy between the values obtained can be explained by the 
use of different density models (the model of density used here is that of Richard, 1999) and by the way the differential equation Eq. \ref {solution_J2n}) is integrated. 
Kosovichev noted that a subsurface shear layer results when the helioseismically obtained 
internal rotation is matched with the surface rotation. Hence, we suspect that the 
subsurface rotation rate, in this very thin layer (called the {\it leptocline}), may play an 
important role. To check this point, we used the rotation model described in 
(Dikpati {\it et al.}, 2002). This model presents the following characteristics: 
(i) the rotation rate is constant in the radiative interior (core), i.e. 
$\Omega_{o}(r,\theta)=435~nHz$ as in Kosovichev (1996); 
(ii) the location and width of the tachocline are assumed to be independent from latitude; 
(iii) the rotation rate at the top of the tachocline and at the surface are given by 
\begin{eqnarray}\
\Omega(r_{cz},\theta) = \Omega_{eq}(r) + a_2 cos^2\theta + a_4 cos^4\theta
\label{standard_rotation_law}
\end{eqnarray}

\noindent
with $a_2$ = -61~$nHz$ and $a_4$ = -73.5~$nHz$; 
(iv) the radial rotation gradient near the surface is assumed to be constant at a given 
latitude and its latitudinal dependence is given by $\beta(\theta)$, which is a polynomial 
function of $cos\theta$: 
$\beta(\theta)=\beta_{0}+\beta_{3}(cos\theta)^{3}+\beta_{6}(cos\theta)^{6}$. 
The transition between layers (of constant $\partial \Omega/ \partial r$ at a given latitude) 
is constructed using error functions, i.e. 
$\Xi_x(r) = 0.5  \left\{1 + erf[2(r - r_x)/w_x]\right\}$ where $w$ is the width of the transition, 
and $x$ stands either for the tachocline (tac), the convective zone (cz) or the surface (s). 
The final expression of the solar internal rotation law is given by Eqs. 1, 2 and 3 of  Dikpati {\it et al.}, (2002) or, alternatively, Eqs. 1, 2, 4 and 5 of (Corbard 
{\it et al.}, 2002).
\noindent
Values of the parameters $\Omega_{o}$, $a_2$ and $a_4$ lead to $\Omega_{cz}$ = 453.5~$nHz$ (but other values of $a_2$ and $a_4$ are found by other authors -see Lefebvre and Rozelot, 2004); we adopted $w_{tac}= w_{cz}= w_{s}= 0.05~R_{\odot}$, 
$r_{tac}= 0.69~R_{\odot}$, $r_{cz}= 0.71~R_{\odot}$ and $r_{s}= 0.97~R_{\odot}$.
Parameter $\beta_0$ has been taken as 437 $nHz/R_{\odot}$, instead of 891.5 used by Kosovichev(1996), and we used 
$\beta_3$ = -214 and $\beta_3$ = -503 ($nHz/R_{\odot}$)\footnote{Dikpati {\it et al.} (2002) 
alternatively considered $\beta_3$ = 0 and $\beta_3$ = -1445 ($nHz/R_{\odot}$), but computations do not show strong differences in $J_{2}$, only 3 per 1000.}. 
The main advantage of this formalism resides in a modeling of the radial differential rotation 
in the subsurface. The order of magnitude of this outward gradient is 
$-5.7~10^{-16}~m^{-1}s^{-1}$ at the equator. 
In a shell with a thickness of about 0.03 $R_{\odot}$, the radial rotation gradient 
$\partial \Omega /\partial r$ is negative from the equator to about 
50$^\circ$ of latitude, crosses zero around 50$^\circ$ and is positive beyond this latitude. 
Basu {\it et al.} (1999) deduced a similar behaviour from their analysis of the splitting of 
high-degree $f$-modes, but finding a reversal of the radial gradient in a zone above 
0.994 $R_{\odot}$ only.
 
After solving equation Eq. \ref{solution_J2n}, it turns out that

\smallskip
\centerline{$J_2$ = 2.28 ~$10^{-7}$ }

\smallskip
\noindent
This computed value is very sensitive to the surface parameters ($a_2$ and $a_4$, see Eq. \ref{standard_rotation_law})
(latitudinal rotation gradient), and to the latitudinal differential rotation in the convective zone. It can reach a maximum of 2.62 ~$10^{-7}$ and a minimum of 1.80 ~$10^{-7}$, 
that is to say a 15\% variation. The situation is worse for $J_4$ where the excursion of the 
computed values is more than 25\%. We conclude that 
(i) helioseismic rotation rates lead to underestimate values of $J_n$ by comparison to values deduced from the theory of Figures (see next section and Table \ref{momgrav}) and 
(ii) the multipole moments ($J_2$ and $J_4$), seem to be sensitive to the physical mechanisms which act at the near surface. Mecheri {\it et al.} (2004), using the same formalism but a different model of density, reached nearly the same conclusions. 
\begin{figure}
\center
\includegraphics[width=.37\textwidth,height=.27\textheight]{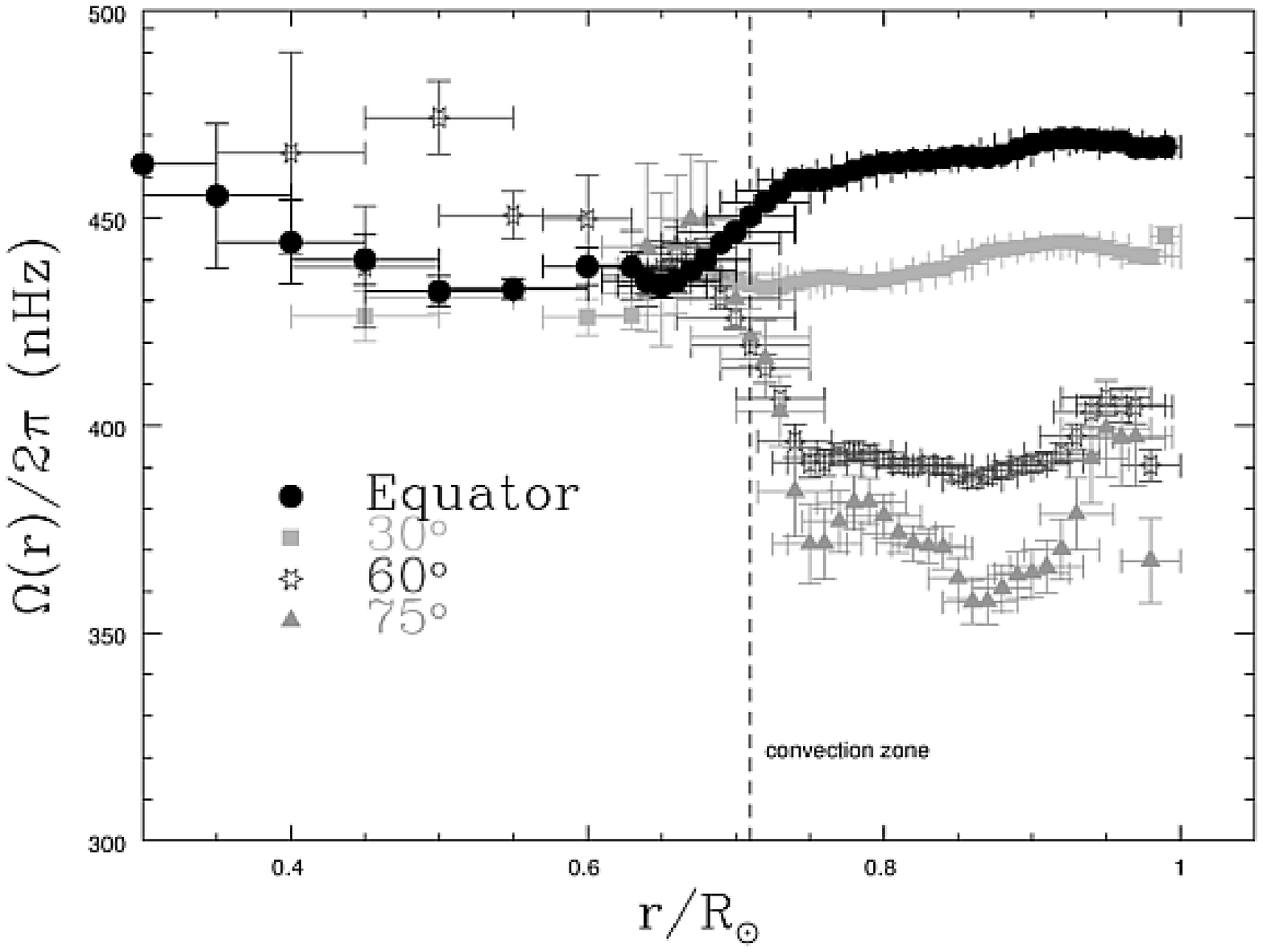}
\includegraphics[width=.25\textwidth, height=.27\textheight]{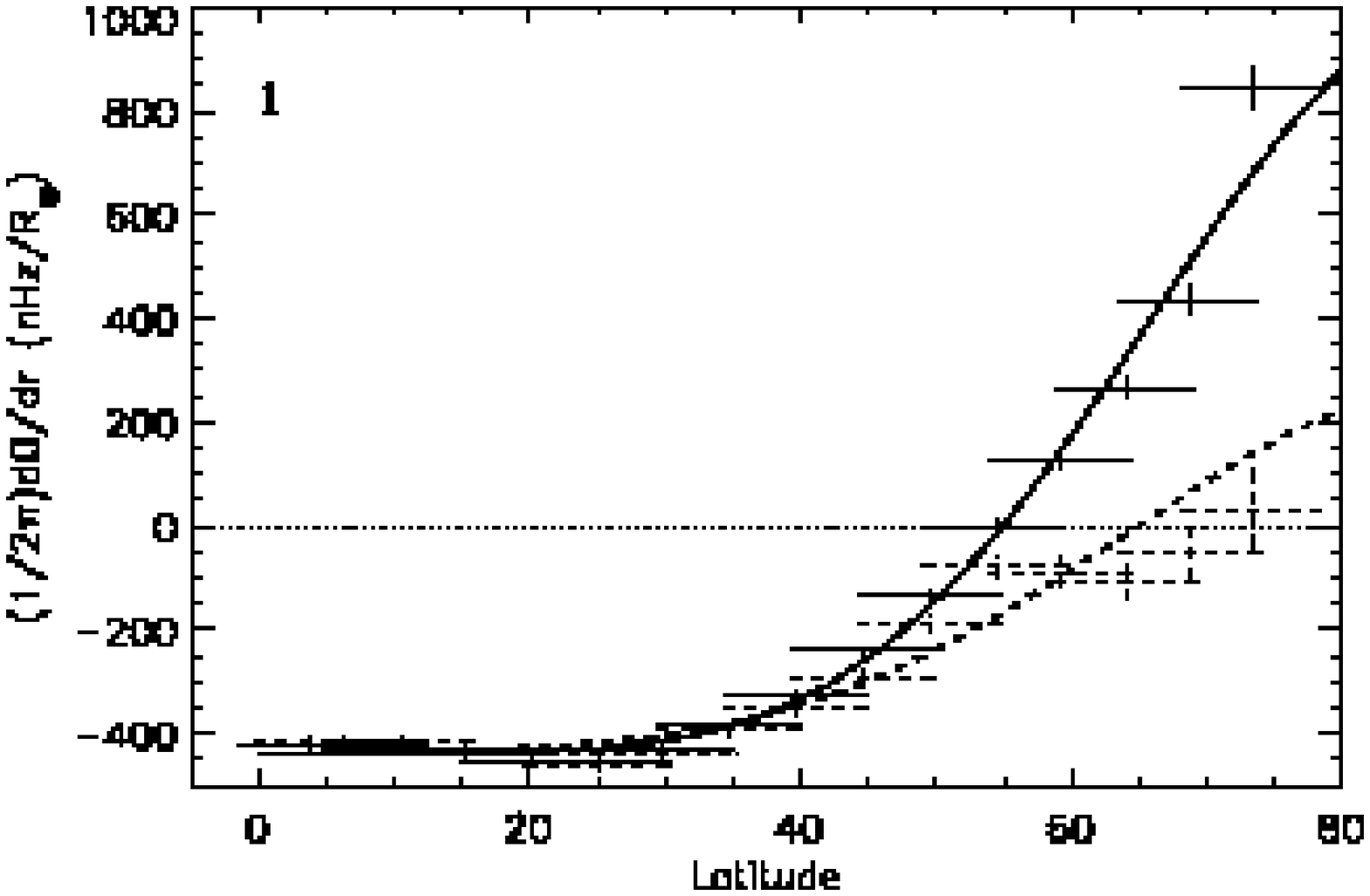}
\includegraphics[width=.34\textwidth, height=.27\textheight]{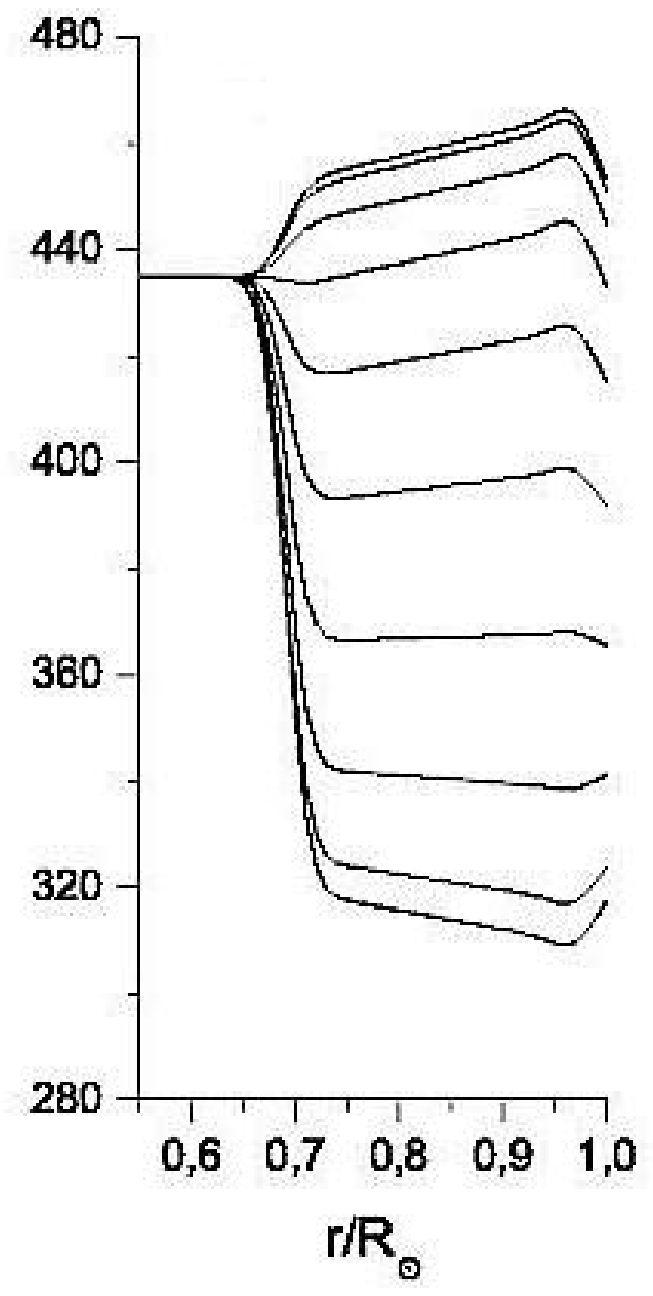}
\caption{\small Left: Solar rotation profiles from $0.3~R_{\odot}$ to
the surface for different latitudes, from the Equator (top curve) to the Pole (bottom curve), 
obtained by a 1.5 dimensional inversion of SOHO-$MDI$ data. 
Courtesy of Maria di Mauro, 2003.
Center: Radial gradient of the angular velocity as a function of latitude. 
See (Corbard {\it et al.}, 2002) for further details.
Right: Solar rotation profiles from $0.55~R_{\odot}$ to the surface for different 
latitudes, computed every $10^{\circ}$ from the Equator (top curve) to the Pole (bottom curve), 
with the analytical rotation model of Corbard {\it et al.} (2002). 
This model takes into account the sub-surface gradient, as shown in the center plot (dotted line).}
\label{corbard}
\end{figure}

\medskip
\noindent
$\bullet$ The second method consists in writing the radius $r$ as a function of the colatitude 
$\theta$,
\begin{equation}
r = R_{sp} \left[ 1 + \sum_{n=1}^{\infty} c_{2n} P_{2n} (cos\theta) \right] \label{eq.rdep} 
\end{equation}
where $R_{sp}$ is the radius of the best fitting sphere passing through the poles and the 
equator of the Sun (for an ellipsoid of revolution with an equatorial radius $a$ and a polar 
radius $b$, $R{_{sp}} = (a^{2} b)^{1/3}$). It is only a matter of algebra to substitute the 
above development (Eq. \ref{eq.rdep}) in the external gravitational potential (Eq. \ref{phi}). 
It is a bit more complicated to determine a centrifugal force in Eq. \ref{clairauttrois} 
corresponding to the rotation law in $(r, \theta)$ and which derives from a potential. 
Adopting the following rotation law fulfills this requirement:
\begin{equation}
        \Omega (r_p) = \Omega_{pol} \left[ 1 + \sum_{i=1}^{\infty} b_{2i} r^{2i}_p 
\sin^{2i}\theta\right]^{1/2}
\label{eq.rotde} 
\end{equation}
where $r_p=r/R_{sp}$ is the normalized radial distance from the solar center at a given depth 
and $\Omega_{pol}$ is the angular velocity at the pole.
Using the solar Greenwich data base, Javaraiah and Rozelot (2002) fitted the parameters:\\
\smallskip
\centerline{ $b_2 = + 0.442$ and $b_4 =  + 0.056$ ~at the surface ($r_p$ = 1); 
$~\Omega_{pol} = 381.8~nHz$} \\
\smallskip
The resulting centrifugal potential can then be expressed on the basis of Legendre polynomials 
and the total potential is written as $\phi_{total}$ = $\Omega_{pol}^2~~ R_{sp}^2~~ h(f, J_n)$ 
where $f$ is the flattening of the Sun ($f = (a-b)/a$ with the above definitions). 
To determine the $J_n$, it is sufficient to consider $\phi_{total}$ as independent from 
$\theta$, that is, the coefficients of $P_{2n}$ vanish. The fact that the analytical differential rotation models lead to a significantly lower 
value of the flatness in comparison to a uniform rotation model can only be interpreted in 
terms of a positive outward rotation gradient in the subsurface.

A full description of the method can be found in Rozelot and Lefebvre (2003) or Lefebvre 
(2003).

\bigskip
The two methods are slightly different. The first one provides the coefficients $J_n$ if the 
mass distribution {\bf and} rotation rate are known. It does not yield a determination of the 
flattening $f$, and there is no means to link $f$ with $J_n$ without strong assumptions. By 
contrast, the second method requires an analytical rotation law (deriving from a potential), 
and it leads to values of $J_n$ expanded in powers of $f$, together with the determination of 
the successive $c_n$. 
}

\section{Results}

{\normalsize
According to the rotation law adopted, the following results are obtained.
\begin{enumerate}
        \item For uniform (rigid) rotation, Chandrasekhar's computations yield a surface 
flatness $f$ that depends on the rotation rate:
\begin{equation}
        f = (0.5 + 0.856 \rho_m / \rho_c ) \alpha
\end{equation}
where $\alpha$ = $\Omega^2$$R_{\odot}$/$g$. The ratio of central to mean density is 
$\rho_c / \rho_m$ and $g$ is the surface gravity. 
For the Sun, one adopts the following values:\\
$\Omega=461.6~nHz$, ~$g=2.7~10^{4}~cm/sec^{2}$, ~$R_{\odot}=6.96~10^{10}~cm$,
$\rho_m=1.409~g/cm^{3}$, $\rho_c=90~g/cm^{3}$.
Accordingly, we find $\alpha$ = 2.17 10$^{-5}$ and  {$f$} = {1.11 $10^{-5}$.}\\
        \item For differential rotation:
                                        \begin{enumerate}
                                        \item The flatness is decreased from 1.11~10$^{-5}$ 
to about 8.85~10$^{-6}$, in contradiction with the theoretical work of Maeder (1999). This can 
be explained only if $\partial \Omega/ \partial r$ is > 0, a behaviour observed at latitudes 
greater than 50$^{\circ}$ (Corbard, 2000; Basu {\it et al.}, 1999). It is  not yet clearly 
established whether this flatness is time-dependent or not.
                                        \item Different values of the successive gravitational 
moments, theoretical or deduced from observations, are given in Table \ref{momgrav}. 
                                        \end{enumerate}
\end{enumerate}
It can be seen that the first method, using helioseimic data, leads to multipole moment values 
lower than those obtained with the second method. However, the octopole moment, $J_4$, is much 
more sensitive than the quadrupole moment, $J_2$, to the presence of latitudinal and radial 
rotations in the convective zone. Taking into account recent values of the subsurface  
rotational gradient, as reported by Corbard and Thompson (2002), the two moments $J_2$ and 
$J_4$ are extremely sensitive to physical mechanisms acting just below the surface. This point 
seems to confirm the existence of a very thin transition layer, as we already suspected, that 
we call the {\it leptocline} (Godier and Rozelot, 2001). }

\begin{center}
\begin{table}[h]
        \begin{tabular}{*{6}{l}} \hline
        \footnotesize References & \footnotesize Method & $ \footnotesize J_{2}$ & $ \footnotesize J_{4}$ & $ \footnotesize J_6$ & \footnotesize Others\\
        \hline
        \hline
                \scriptsize Ulrich \& Hawkins  &  \scriptsize SSE + & \scriptsize (10-15)~10$^{-8}$ & \scriptsize (0.2-0.5)~10$^{-8}$  & &  \\
         \scriptsize ~~~~ (1981)& \scriptsize spots rotation law& \scriptsize & \scriptsize  &     \\
        \hline
        \scriptsize Gough (1982)  &  \scriptsize First determination of &  \scriptsize 36~10$^{-7}$  &  &  &     \\
         \scriptsize & \scriptsize helioseismic rot. rates &  & & &     \\
        \hline
        \scriptsize Campbell \& & \scriptsize Planetary orbits  & \scriptsize (5.5 $\pm$ 1.3)~10$^{-6} $  &  &  & \\
         \scriptsize Moffat (1983)&  &  & & &     \\
        \hline
        \scriptsize Landgraf    &  \scriptsize Astrometry of&  \scriptsize (0.6 $\pm$ 5.8)~10$^{-6}$ &  &  &     \\
          \scriptsize ~~~~(1992) &   \scriptsize minor planets & & &     \\
        \hline
        \scriptsize Lydon \& Sofia   & \scriptsize SDS & \scriptsize 1.84~10$^{-7}$ & \scriptsize 9.83~10$^{-7}$& \scriptsize 4~10$^{-8}$ & \scriptsize $J_8$ = -4~10$^{-9}$  \\
        \scriptsize ~~~~ (1996)& \scriptsize Experiment & & & &  \scriptsize $J_{10}$ = -2~10$^{-10}$   \\
        \hline
        \scriptsize Patern\`o {\it et al.} (1996) & \scriptsize SSE + empirical& \scriptsize 2.22~10$^{-7}$& & &     \\
        & \scriptsize rotation law and SDS & & & &     \\
        \hline
        \scriptsize Pijpers (1998) &\scriptsize SSE  + GONG& \scriptsize (2.14$\pm$0.09) ~10$^{-7}$  & &     \\
                 & \scriptsize and SOI/MIDI data& \scriptsize (2.23$\pm$0.09) ~10$^{-7}$  & &     \\
        & \scriptsize Weighted value & \scriptsize (2.18$\pm$0.06) ~10$^{-7}$  & &     \\
        \hline
        \scriptsize Armstrong \& & \scriptsize Vect. Spher. Harm.& \scriptsize -0.222~10$^{-6}$ &\scriptsize 3.84~10$^{-9}$& &   \\
        \scriptsize Kuhn (1999)& \scriptsize numerical error& \scriptsize 0.002 ~10$^{-6}$& \scriptsize 0.4~10$^{-9}$& &    \\
        \hline
        \scriptsize Godier \&  & \scriptsize SSE +  & \scriptsize 1.6~10$^{-7}$ &  &  &     \\
         \scriptsize Rozelot (1999)& \scriptsize Kosovichev law &  & & &     \\
        \hline
        \scriptsize Roxburgh (2001) & \scriptsize SSE + 2 models of  & \scriptsize 2.208~10$^{-7}$ & \scriptsize -4.46~10$^{-9}$ & \scriptsize -2.80~10$^{-10}$  &  \scriptsize $J_8$ = 1.49~10$^{-11}$ \\
         & \scriptsize rotation law & \scriptsize 2.206~10$^{-7}$ & \scriptsize -4.44~10$^{-9}$ & \scriptsize -2.79~10$^{-10}$ & \scriptsize $J_8$ = 1.48~10$^{-11}$ \\
         \hline
        \scriptsize Rozelot {\it et al.} (2001)& \scriptsize Theory of Figures & \scriptsize -(6.13 $\pm$ 2.52)~10$^{-7}$ & \scriptsize 3.4~10$^{-7}$  &  &    \\
        \scriptsize  & & \scriptsize Note 3 &  \scriptsize Note 4   &  &    \\
        \hline
        \scriptsize Rozelot \& Lefebvre & \scriptsize Theory of Figures & \scriptsize - 6.52~10$^{-7} $& \scriptsize 4.20~10$^{-7}$  & \scriptsize -9.46~10$^{-9}$ & \scriptsize $J_8$ = 2.94~10$^{-13}$   \\
        \scriptsize ~~~~ (2003)&  & & & &  \\
        \hline
        \scriptsize Present paper & \scriptsize SSE + & \scriptsize - 2.28~10$^{-7}$ $\pm$ 15 $\%$ & \scriptsize  very sensitive &  &   \\
        &   \scriptsize Subsurface gradient&  & \scriptsize to SGR & &  \\
        &   \scriptsize of rotation (SGR)&  & \scriptsize range: $\pm$ 20\% & &  \\
        \hline\hline
        \multicolumn{6}{l}{\scriptsize Note 1: SDS stands for Solar Disk Sextant} \\
        \multicolumn{6}{l}{\scriptsize Note 2: SSE stands for Stellar Structure Equations (based on Eqs. \ref{clairauttrois} and \ref{poisson}).} \\
        \multicolumn{6}{l}{\scriptsize Note 3: The apparent large error comes from the fact that the value is a weighted average of several rotation rates.} \\
        \multicolumn{6}{l}{\scriptsize Note 4: A mistake has been made in the second term of $J_4$ (i.e. $mA_4$),
        which was incorrectly multiplied} \\
        \multicolumn{6}{l}{\scriptsize  ~~~~~~~~~~ by ``$f$'' in the computations.} \\
        \end{tabular}
        \caption{\small Solar gravitational multipole moments quoted from different authors and methods.      } 
        \label{momgrav}
\end{table}
\end{center}

\section{Relevance of precise solar gravitational moments}

{\normalsize The determination of the outer shape of the Sun has at least four major astrophysical 
consequences:

\medskip
\noindent
{\bf 1.} {\it Accurate determination of the successive gravitational multipole moments}, and 
$J_2$ in particular. Until recently, $J_2$ was not really known, values ranging from 
1.0~10$^{-7}$ (Ulrich and Hawkings, 1981) to 36~10$^{-7}$ (Gough, 1982). However, on one hand, 
it is expected that the flattening $f$ be lower than Roche's limit, which is, in the solar 
case 2.7~10$^{-5}$. This means that $J_2$ must be lower than 1.18~10$^{-5}$.
On another hand, we were able to fix an upper bound to the quadrupolar moment, 
$J_{2~max}$ = 3.0~10$^{-6}$, meaning $f$ is lower than 1.38~10$^{-5}$ (Rozelot and Bois, 1998). 
Beyond this value, there would be observational incompatibilities for specific astrophysical 
phenomena such as lunar librations.
\newline
Helioseismic data always lead to lower values of the calculated parameters. That is the case 
for instance for the photospheric radius (6.95508~10$^{10}$ $cm$ instead of 
6.9599~10$^{10}$ $cm$), for the flattening itself (6~10$^{-6}$ instead of 9~10$^{-6}$), for the 
velocity rate at the equator (452.3 $nHz$ instead of 467.3 $nHz$ (Javaraiah, 1995\footnote{solar rotation rate deduced from the sunspots is greater than 461 nHz, see for instance Table 3.2 in Lefebvre, 2003.}, and also 
for $J_2$. In this context, Pireaux and Rozelot (2003) adopted the following theoretical 
range for the solar quadrupole moment, based on helioseismic data (Patern\`{o} {\it et al.}, 1996; 
Pijpers, 1998) and on a solar stratified model taking into account latitudinal differential 
rotation (Godier and Rozelot, 1999),

\begin{equation}
      J_2 = (2.0 \pm 0.4)~ 10^{-7}
\label{adopted_J2} 
\end{equation}
whereas the theory of Figures (Lefebvre, 2003) yields 
\begin{equation}
		  J_2 = (6.52 \pm 2.55)~ 10^{-7}
\label{theory_figure_J2} 
\end{equation}
These discrepancies still need to be resolved.

\smallskip
\noindent
{\bf 2.} {\it Accurate determination of the perihelion advance of a planet}, 
for which a knowledge of the solar quadrupole moment would help to determine the relativistic 
post-Newtonian parameters $\gamma$ and $\beta$. 
Indeed, the perihelion advance (with respect to the classical Keplerian prediction) of planets 
($\Delta W$) is a combination of a purely relativistic effect and a contribution from the Solar 
quadrupole moment. It is given by the following expression\footnote{In some references, the 
coefficient of the term containing the contribution of the orbital inclination is improperly 
written.}: 
 
\begin{equation}
        \Delta W = \Delta W_{0,GR} \left[\frac{1}{3}(2\alpha^2+2\alpha\gamma-\beta) - 
        \mbox{\slshape F($\alpha$)} J_2 \right]
\end{equation}
where $\Delta W_{0,GR}$ represents the contribution due to General Relativity with $J_{2}$ 
set to zero; $\alpha$, $\beta$, $\gamma$, the parameters describing the theory of the 
gravitation in the parameterized post-Newtonian formalism; $\cal F(\alpha)$ is a coefficient 
depending on the considered planet (see Pireaux and Rozelot, 2003) for further developments.
\newline
In General Relativity, $\alpha =\beta =\gamma = 1$, so that the perihelion advance of planets 
can be written as:  
$\Delta W$ = 42.981 [ 1 - $\cal F$ $J_2$]  (arcsec/cy), 
with $\cal F$ = [$R_{\odot}^2c^2(3 sin^2 i -1)$]/[$2G M_{\odot} a (1-e^2)$] where $G$, $c$, 
$R_{\odot}$ and $M_{\odot}$ have their usual meaning; and $a$, $e$ and $i$ are respectively 
the semi-major axis, the eccentricity and the inclination of the orbit of the body in question. 
The contribution of $J_2$ to $\Delta W$ is relevant. 
In the case of Mercury, $\cal F$= +2.8218 $10^{3}$.  

The parameter $\gamma$ is constrained by the CASSINI doppler experiment 
(Bertotti {\it et al.}, 2003) to
\begin{equation}
\gamma -1 =(-2.1\pm 2.3)~10^{-5}  
\label{CASSINI_data}
\end{equation}
while observational constraints on the Nordtvedt effect (Williams {\it et al.}, 2001) , 
combined with the observational range for $\gamma$, leads to 
\begin{equation}
\beta \in \left[ 0.9998;1.0003\right]
\label{llr_CASSINI_data} 
\end{equation}
Hence, it turns out that General Relativity is not excluded by those Solar System experiments. 
However, General Relativity would be incompatible with the Mercury perihelion advance test if 
$J_2=0$ was assumed. But with the adopted theoretical range for $J_2$  in Eq. \ref{adopted_J2}, 
General Relativity agrees with this latter test... and there is still room for an alternative 
theory too (see Fig. \ref{ppn}).   

\begin{figure}[htb]
\center
\includegraphics[width=.325\textwidth]{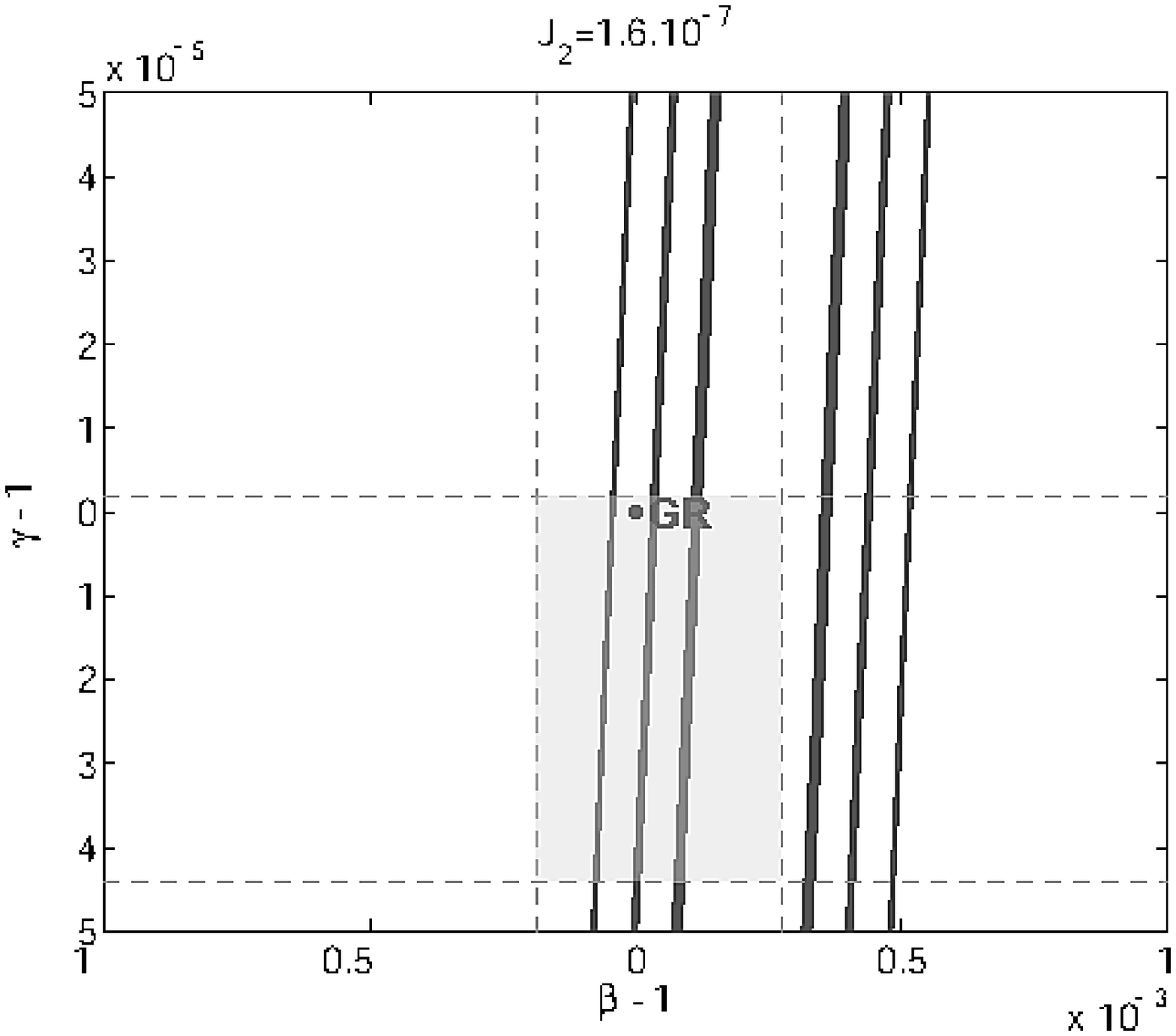}
\includegraphics[width=.325\textwidth]{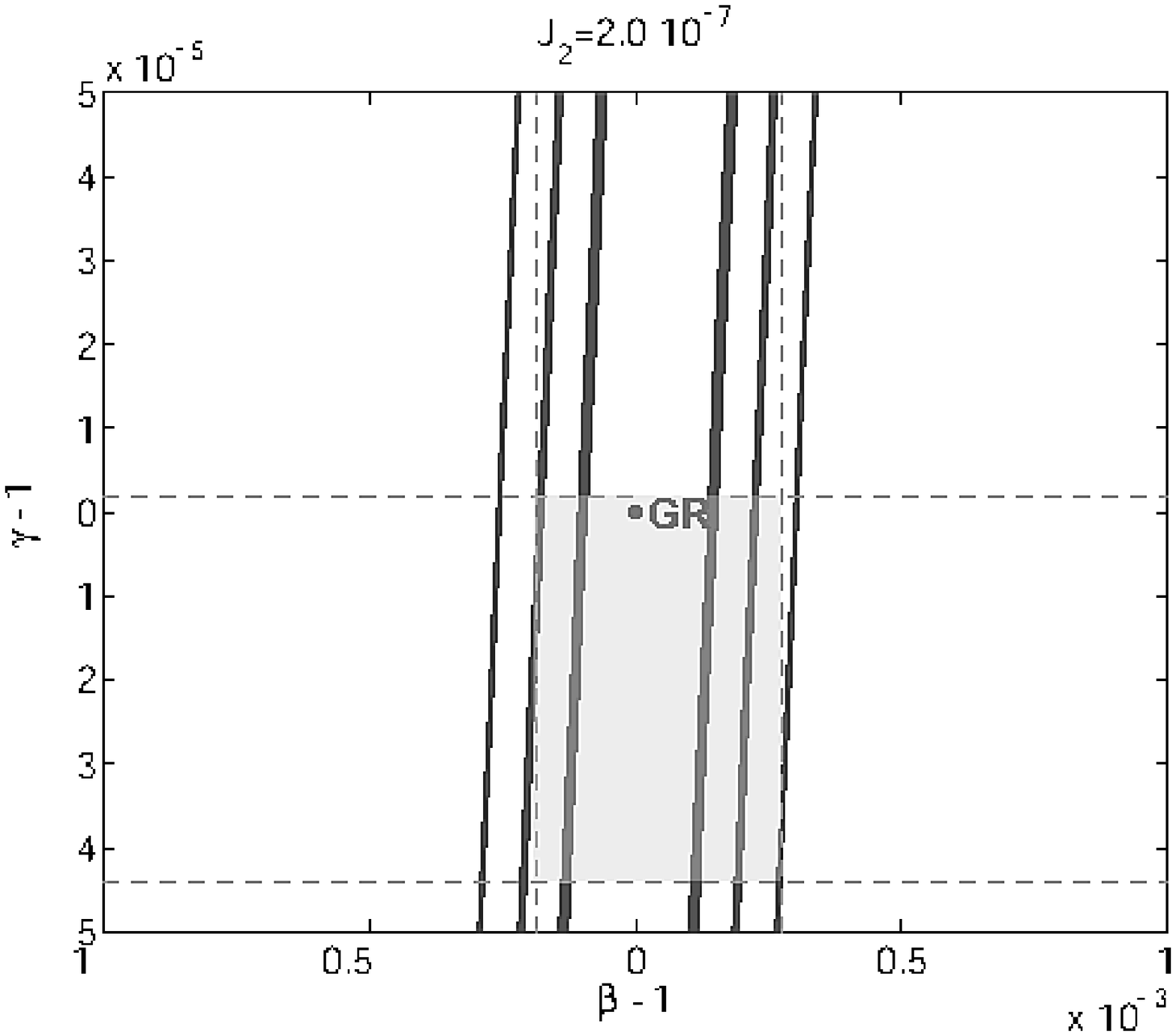}
\includegraphics[width=.325\textwidth]{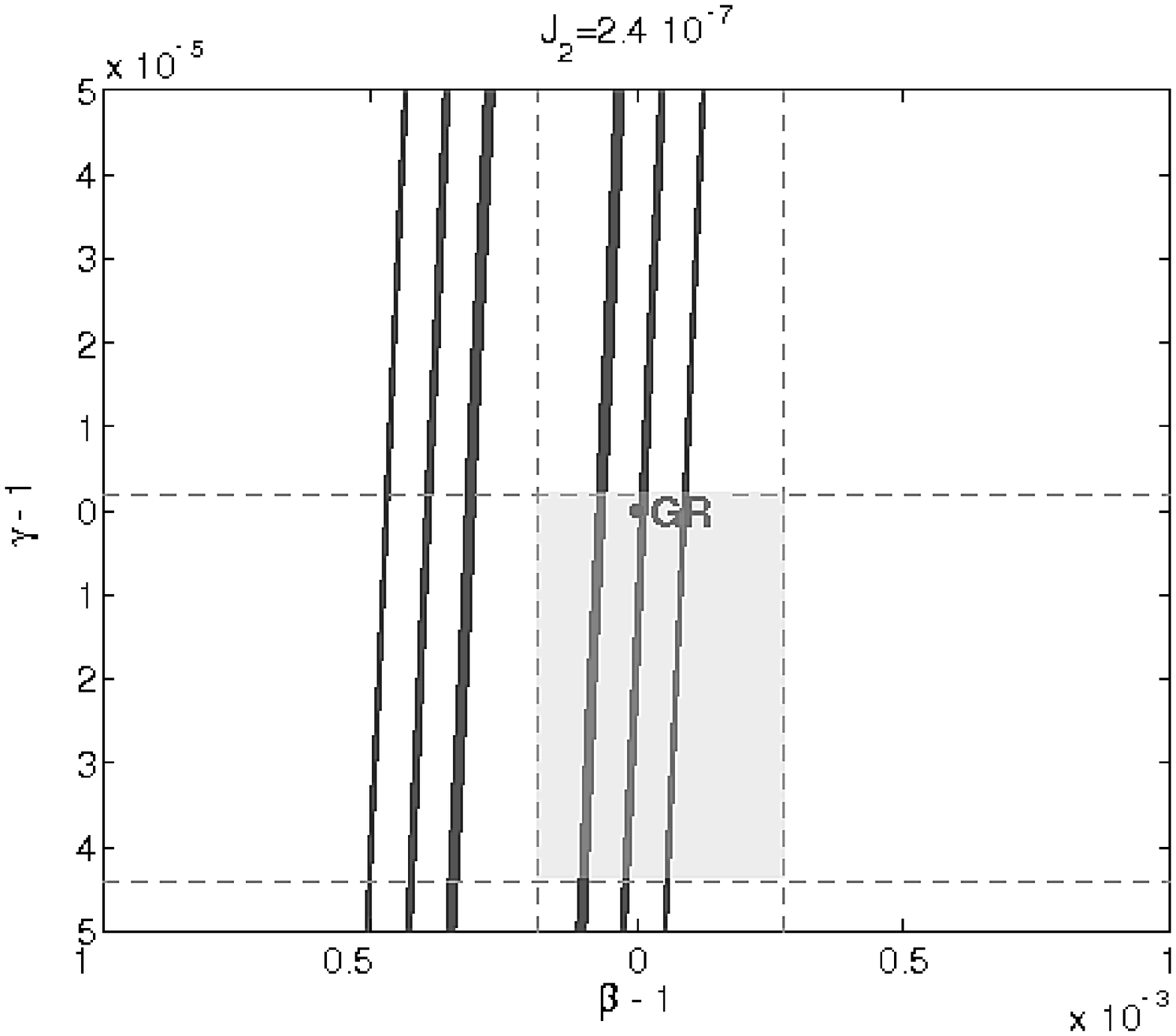}
        \caption{\small For a given value of $J_2$ the perihelion advance of Mercury constitutes 
a test of the PN parameters $\beta$ and $\gamma$.\newline        
In the $\beta$ (Eq. \ref{llr_CASSINI_data}) and $\gamma$ (Eq. \ref{CASSINI_data}) plane 
($\alpha $ set to 1), we have plotted 1$\sigma$ (the smallest), 2$\sigma$ and 3$\sigma$ 
(the largest) confidence level ellipses. Those are based on the values for the observed 
perihelion advance of Mercury, $\Delta W_{obs}$, given in the literature and summarized in 
reference (Pireaux and Rozelot, 2003).  
Remark also that the position of the ellipses varies according to the value of $J_{2}$ chosen; 
but, their orientation is determined by the combination ($2\alpha ^{2}+2\alpha \gamma -\beta $) 
that appears in the expression for $\Delta W$. Nevertheless, General Relativity is still in 
the 3$\sigma$ contours for the allowed theoretical values of $J_{2}$ argued by the authors 
(Eq. \ref{adopted_J2}). 
Fig. 1 a, b, c represent the confidence contours for $\beta$, and $\gamma$, 
$J_{2}$ fixed to its minimum, average and maximum value respectively (Eq. \ref{adopted_J2}).}
        \label{ppn}
\end{figure}

\medskip
\noindent
{\bf 3.} {\it Accurate determination of the precession of the orbital plane of a planet 
about the Sun's polar axis.} $J_2$ also contributes to this relativistic effect, which is 
unaffected by the relativistic PN combination ($2\alpha ^{2}+2\alpha \gamma -\beta $). 
The precession of the orbital plane is more easily discernible for moderately large values 
of the inclination $i$.

\medskip
\noindent
{\bf 4.} {\it Modern planetary ephemerides} now include a non null value of $J_2$. However, 
they are presently not able to infer simultaneously all the usual parameters 
(masses, radii,..., PN parameters) and $J_2$ from fits to observational data, due to strong 
correlations. The order of magnitude adopted for $J_2$ corresponds to $10^{-7}$, 
but, for example, the estimation of the PN parameters is rather tolerant to the assumed value 
of $J_2$.

\section{Observations }

{\normalsize Measurements of $J_n$ are not directly accessible.
\newline
The multipole moments could be measured dynamically by sending and accurately tracking a 
probe carrying a drag-free guidance system to within a few radii of the solar center. The 
$J_n$ are then inferred from the precise determination of the trajectory (Pireaux and 
Rozelot, 2003).
\newline
Alternatively, the $J_n$ can be inferred from in orbit measurement of solar properties 
(Rozelot and Godier, 2002). Indeed, solar asphericities, encoded mainly by the first two 
coefficients $c_2$ and $c_4$ (Eq. \ref{eq.rdep}) can be observed. An estimate of these two 
coefficients has been recently derived from SOHO-$MDI$ space-based observations 
(Armstrong and Kuhn, 1999):

\smallskip
\centerline{$c_2$ = (-5.27 $\pm$ 0.38)~10$^{-6}$ and $c_4$ = (+1.3 $\pm$ 0.51)~10$^{-6}$. }
\smallskip

\noindent
These results were obtained by measuring small displacements of the solar-limb darkening 
function (for further details, see Kuhn {\it et al.}, 1998), and the $c_n$ coefficients are 
comparable to an isodensity surface level (see footnote 1).
>From Earth-based observations at the scanning heliometer of the Pic du Midi Observatory, we 
also obtained estimates of $c_2$ and $c_4$ coefficients. We have shown that asphericities can 
be observed. A bulge appears to extend from the equator to the edge of the royal zones (about 
50$^{\circ}$ latitude), with a depression beyond (Rozelot {\it et al.}, 2003). Such a 
distorted shape can be interpreted through the combination of the quadrupole and hexadecapole 
terms, which, as shown previously, directly reflect the non uniform velocity rate in surface. 
Moreover, this distribution implies a thermal wind effect, blowing in the low density surface 
layers of the Sun, from the poles towards the equator, just as in terrestrial meteorology. 
This phenomenon was recently studied by  Lefebvre and Rozelot (2004) who explained for the 
first time the existence of small distortions, not exceeding some 20 mas of amplitude. 
Fig. \ref{pic2003} illustrates those observations. Accordingly, We found:

\smallskip
\centerline{Year 2000: $c_2$ = (-7.6 $\pm$ 0.2)~10$^{-6}$ and $c_4$ = +2.2 ~10$^{-6}$ }
\centerline{Year 2001: $c_2$ = (-1.1 $\pm$ 0.5)~10$^{-5}$ and $c_4$ = +3.4 ~10$^{-6}$ }
\centerline{Year 2002: $c_2$ = (-3.8 $\pm$ 0.8)~10$^{-5}$ and $c_4$ = +2.5 ~10$^{-6}$ }
\smallskip

\noindent
The differences in the estimates stem from the difficulty of observations, mainly due to 
seeing conditions. The mean observed value of $c_2$, -7.5~10$^{-6}$, is not too far from 
the theoretical one, $\approx$ -(2/3)$f$$\approx$-5.9~10$^{-6}$ with $f$ = 8.9~10$^{-6}$ based 
on a solar model with a differential rotation law. 
The coefficient $c_4$ remains difficult to match with the theory, which predicts +(12/35)$f^2$ 
for a uniform rotation law. The only explanation is that the distorted shape coefficient $c_4$ 
is very sensitive to surface phenomena.

\begin{figure}
\center
\includegraphics[width=.49\textwidth,height=0.23\textheight]{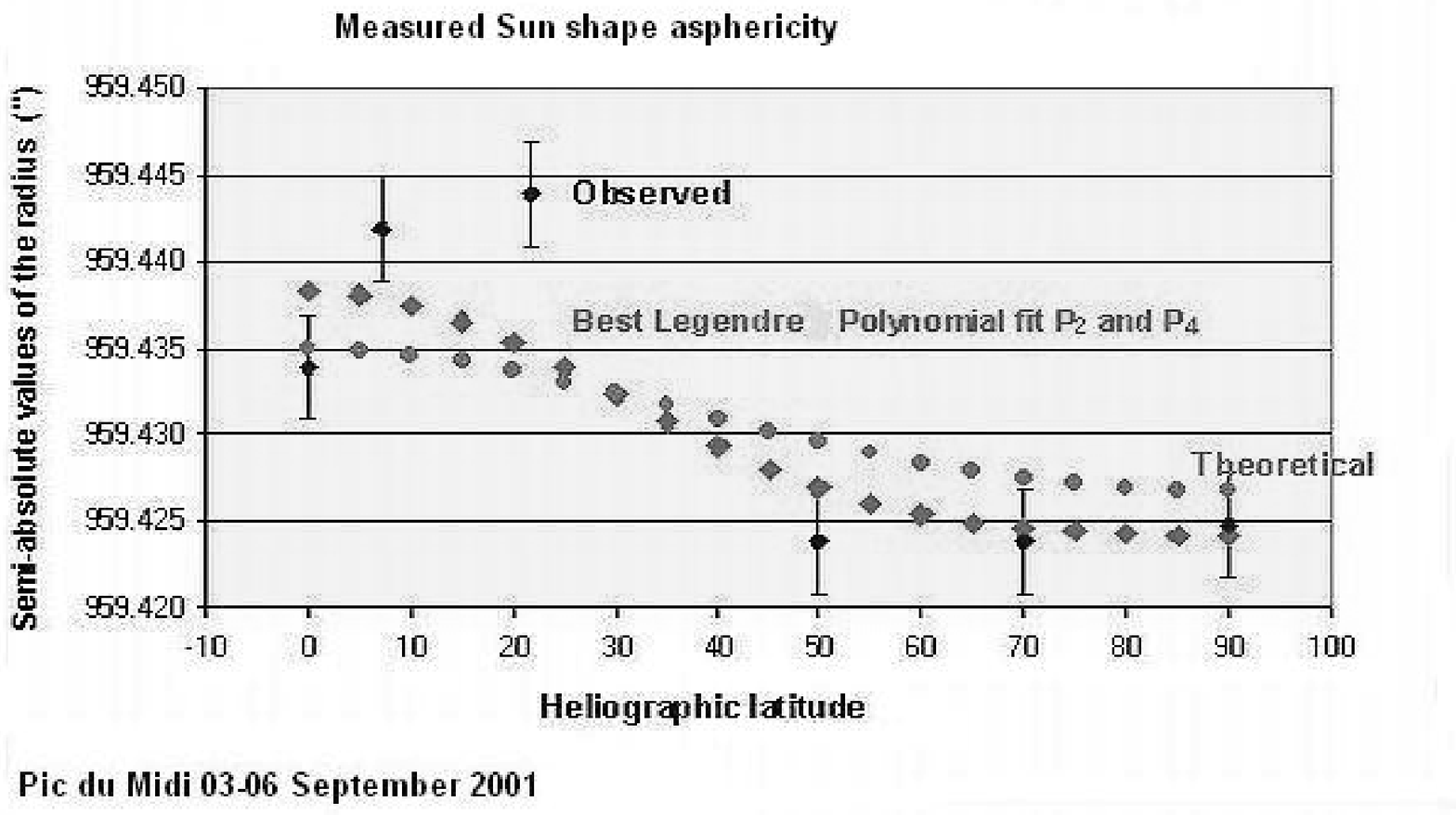}
\includegraphics[width=.49\textwidth, height=0.23\textheight]{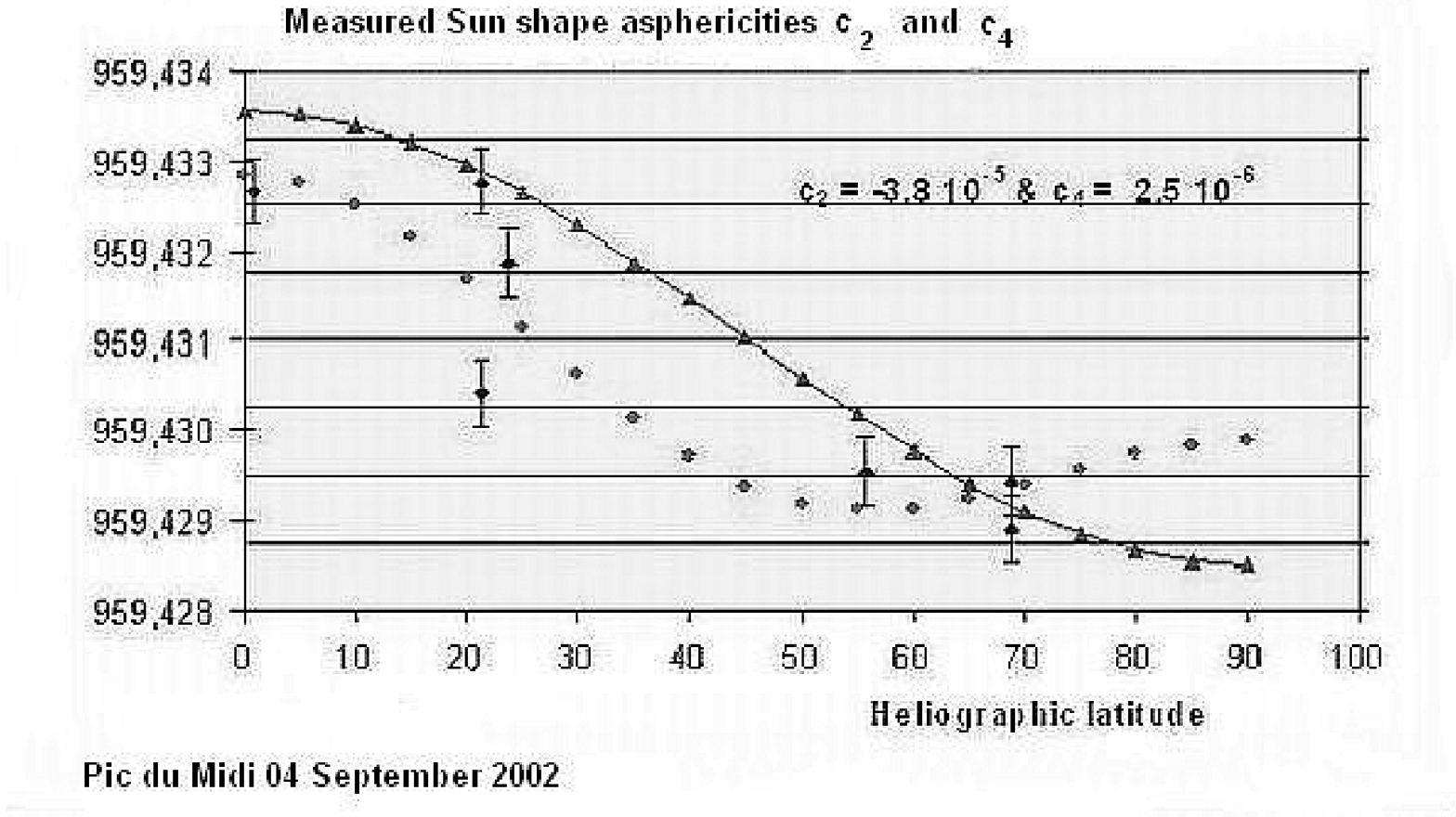}
        \caption{\small Observations of the solar asphericities by means of the scanning 
heliometer at the Pic du Midi Observatory, in 2001 (left) and in 2002 (right). 
On the left plot, $c_2$ and $c_4$ derive directly from Eq. \ref{eq.rdep}, whereas, on the 
right plot, these $c_n$ come from a fit to the best ellipsoid passing through the measured 
points. }
        \label{pic2003}
\end{figure}
}

\section{Conclusions}

{\normalsize The main conclusions that can be drawn from this review are the following:

1. Helioseismic data, even when reliable, imply a solar surface rotation rate in conflict with 
other observed photospheric rotation data. As a consequence, the values of both the solar 
gravitational moments ($J_n$) and the shape coefficients ($c_n$) are slightly different, 
larger in the latter case. These discrepancies still need to be resolved.

2. The latitudinal dependence conveys sub-surface physical mechanisms that can be explained 
theoretically. Thus, in spite of the fact that ground-based observations are altered by 
seeing effects amplifying or superimposing noise, it could be suggested that the solar shape 
is not a pure spheroid. It is clear that improvements on this question will depend on better 
understanding of the sub-surface layers and on better measurements of the limb shape. 

3. The exact shape of the Sun critically depends on the rotation law in the external layers. 

4. In the setting of General Relativity, it is clear that accurate measurements of the 
perihelion advance of small planets such as Icarus will help to better determine the 
quadrupole moment (dynamical estimate). Measurements of the multipole moments ($J_n$) will be 
obtained as a by-product of two dictinct space missions: Beppi-Colombo (scheduled for 2009) 
and GAIA (scheduled for 2010). These missions are essential for future developments on that 
question. 

5. In-orbit solar measurements with sufficient resolution are also essential. Measurements of 
the asphericities ($c_n$) and determination of the true helioid are one of the main goals of 
the future space mission PICARD, scheduled to be launched by 2008; and also the main targets 
of balloon flights, such as SDS flights.}

{\bf Aknowledgements.} We thank Thierry Corbard for reviewing the manuscript and his helpful comments.

\end{document}